\documentclass[epj]{svjour}

\def\be{\begin{equation}} 
\def\ee{\end{equation}}   

\RequirePackage{graphicx}

\usepackage{cite} 
\usepackage{amsmath}
\usepackage{hyperref}
\usepackage{cite}
\usepackage{amsmath,amssymb}
\usepackage{xcolor}
\usepackage[utf8]{inputenc}

\usepackage{graphics}

\usepackage{cite} 
\usepackage{amsmath}
\usepackage{hyperref}
\usepackage{cite}
\usepackage{amsmath,amssymb}
\begin{document}
\title{Interior solutions of relativistic stars with anisotropic matter 
in scale-dependent gravity} 


\author{
Grigoris Panotopoulos \inst{1} 
\thanks{E-mail: \href{mailto:grigorios.panotopoulos@tecnico.ulisboa.pt}{\nolinkurl{grigorios.panotopoulos@tecnico.ulisboa.pt}} }
\and
\'Angel Rinc\'on \inst{2}
\thanks{E-mail: \href{mailto:angel.rincon@pucv.cl}{\nolinkurl{angel.rincon@pucv.cl}} }
\and
Il{\'i}dio Lopes \inst{1} 
\thanks{E-mail: \href{mailto:ilidio.lopes@tecnico.ulisboa.pt}{\nolinkurl{ilidio.lopes@tecnico.ulisboa.pt}} }
}
%
%
\institute{ 
Centro de Astrof\'{\i}sica e Gravita{\c c}{\~a}o-CENTRA, Departamento de F{\'i}sica, Instituto Superior T\'ecnico-IST, \\
Universidade de Lisboa-UL, Av. Rovisco Pais, 1049-001 Lisboa, Portugal.
\and
Instituto de F{\'i}sica, Pontificia Universidad Cat{\'o}lica de Valpara{\'i}so, Avenida Brasil 2950, Casilla 4059, Valpara{\'i}so, Chile.
}
\date{Received: date / Revised version: date}
%
\abstract{
We obtain well behaved interior solutions describing hydrostatic equilibrium of anisotropic relativistic stars in scale-dependent gravity, where Newton's constant is allowed to vary with the radial coordinate throughout the star. Assuming i) a linear equation-of-state in the MIT bag model for quark matter, and ii) a certain profile for the energy density, we integrate numerically the generalized structure equations, and we compute the basic properties of the strange quark stars, such as mass, radius and compactness. Finally, we demonstrate that stability criteria as well as the energy conditions are fulfilled. Our results show that a decreasing Newton's constant throughout the objects leads to slightly more massive and more compact stars.
\PACS{
      {PACS-key}{discribing text of that key}   \and
      {PACS-key}{discribing text of that key}
     } 
} 
\maketitle
%

\section{Introduction}\label{Intro}

Einstein's theory of General Relativity (GR) \cite{GR} is a relativistic theory of gravitation, which not only is beautiful but also very successful \cite{tests1,tests2}. The classical tests and solar system tests \cite{tests3}, and recently the direct detection of gravitational waves by the LIGO/VIRGO collaborations \cite{Ligo} have confirmed a series of remarkable predictions of GR.

\smallskip

Despite its success, however, it has been known for a long time that GR is a classical, non-renormalizable theory of gravitation. Formulating a theory of gravity that incorporates quantum mechanics in a consistent way is still one of the major challenges in modern theoretical physics. All current approaches to the problem found in the literature (for a partial list see e.g. \cite{QG1,QG2,QG3,QG4,QG5,QG6,QG7,QG8,QG9} and references therein), have one property in particular in common, i.e. the basic quantities that enter into the action describing the model at hand, such as Newton's constant, gauge couplings, the cosmological constant etc, become scale dependent (SD) quantities. This of course is not big news, as it is known that a generic feature in ordinary quantum field theory is the scale dependence at the level of the effective action. 

\smallskip

As far as black hole physics is concerned, the impact of the SD scenario on properties of black holes, such as thermodynamics or quasinormal spectra, has been studied over the last years, and it has been found that SD modifies the horizon, thermodynamic properties and the quasinormal frequencies of classical black hole backgrounds \cite{SD1,SD2,SD3,SD4,SD5,SD6,SD7}. Moreover, a scale dependent gravitational coupling is expected to have significant cosmological and astrophysical implications as well. In particular, since compact objects are characterized by ultra dense matter and strong gravitational fields, a fully relativistic treatment is required. Naturally, it would be interesting to investigate the impact of the SD scenario on properties of relativistic stars. 

\smallskip

In the present work we propose to obtain for the first time interior solutions of relativistic stars with anisotropic matter in the SD scenario, extending a previous work of ours where we studied isotropic compact objects \cite{PRL2}. In particular, here we shall focus on strange quark stars, which comprise a less conventional class of compact stars. Although as of today they remain hypothetical astronomical objects, strange quarks stars cannot conclusively be ruled out yet. As a matter of fact, there are some claims in the literature that there are currently some observed compact objects exhibiting peculiar features (such as small radii for instance) that cannot be explained by the usual hadronic equations-of-state used in neutron star studies, see e.g. \cite{cand1,cand2,cand3}, and also Table 5 of \cite{weber} and references therein. The present study is also relevant for the possible implications to understand the nature of compact stars. Recently, a few authors suggested that strange matter could exist in the core of NS-hybrid stars \cite{Benic:2014jia,2019arXiv190204887Y,2019EPJC...79..815E}, while others claim such stars are almost indistinguishable from NS \cite{Jaikumar}.

\smallskip

Celestial bodies are not always made of isotropic matter, since relativistic particle interactions in a very dense nuclear matter medium could lead to the formation of anisotropies \cite{paper1}. The investigation of properties of anisotropic relativistic stars has received a boost by the subsequent work of \cite{paper2}. Indeed, anisotropies can arise in many scenarios  of a dense matter medium, such as phase transitions \cite{paper3}, pion condensation \cite{paper4}, or in presence of type $3A$ super-fluid \cite{paper5}. See also \cite{anisotropia,Gabbanelli,PRL1,Tello-Ortiz} as well as \cite{Ref_Extra_1,Ref_Extra_2,Ref_Extra_3} for more recent works on the topic, and references therein. In the latter works relativistic models of anisotropic quark stars were studied, and the energy conditions were shown to be fulfilled. In particular, in \cite{Ref_Extra_1} an exact analytical solution was obtained, in \cite{Ref_Extra_2} an attempt was made to find a solution to Einstein's field equations free of singularities, and in \cite{Ref_Extra_3} the Homotopy Perturbation Method was employed, which is a tool that facilitates working with Einstein's field equations.

\smallskip

Currently there is a rich literature on relativistic stars, which indicates that it is an active and interesting field.
For stars with a net electric charge see e.g. \cite{1995MNRAS.277L..17D,1999CQGra..16.2669D,1982Ap&SS..88...81Z,2003PhRvD..68h4004R,2007BrJPh..37..609S,2013PhRvD..88h4023A,2014PhRvD..89j4054A,2009PhRvD..80h3006N,2015PhRvD..92h4009A}, for stars with anisotropic matter see e.g. \cite{Sharma:2007hc,2002ChJAA...2..248M,2017AnPhy.387..239D,Deb:2015vda,Bhar:2016fne,Gabbanelli:2018bhs,Maurya:2015qfm,Deb:2016lvi,Chowdhury:2019gbq,Chowdhury:2019wte}, and for charged ansotropic objects see e.g. \cite{Thirukkanesh:2008xc,Varela:2010mf,Maurya_v1,Deb:2018ccw,Morales:2018nmq,2019EPJC...79...33M,Singh:2018nss,Bhar:2017hbw}, and also compact stars with specific mass function \cite{Maurya:2017aop}.

\smallskip

The plan of our work is the following: In the next section we briefly review the SD scenario. After that, in Sect. \ref{Hydro} we present the generalized structure equations that describe hydrostatic equilibrium of relativistic stars.
Then, in the fourth section we introduce the equation-of-state, we obtain the interior solutions integrating the structure equations numerically, and we also show that the solutions obtained here are realistic, well behaved solutions. Finally, we summarize our work and finish with some concluding remarks in the final section. We adopt metric signature, $(-,+,+,+)$, 
and we work in units where the speed of light in vacuum, $c$, and the usual Newton's constant, $G_N$, are set to unity.

\section{Scale-dependent gravity}

The aim of this section is to briefly introduce the formalism. 

The asymptotically safe gravity program is one of the variety of approach of quantum gravity an this is, precisely, the inspiration of our formalism. Also, close-related approaches share similar foundations, for instance the well-known Renormalization group improvement method \cite{Bonanno:2000ep,Bagnuls:2000ae,Bonanno:2001xi,Reuter:2003ca} (usually applied to black hole physics) or the running vacuum approach \cite{Sola:2016jky,Sola:2013gha,Shapiro:2009dh,Basilakos:2009wi,
Hernandez-Arboleda:2018qdo,Panotopoulos:2019xbw} (usually implemented in cosmological models).
Following the same philosophy, recently scale-dependent gravity has provided us with non-trivial black holes solutions as well as cosmological solutions, investigating different conceptual aspects and offering novel results (see, for instance 
\cite{Koch:2016uso,Rincon:2017ypd,Rincon:2017goj,Rincon:2017ayr,
Contreras:2017eza,Rincon:2018sgd,Contreras:2018dhs,Rincon:2018lyd,
Rincon:2018dsq,Contreras:2018gct,Rincon:2019cix,Rincon:2019zxk,
Contreras:2019fwu,Fathi:2019jid,Panotopoulos:2019qjk,
Contreras:2019nih,Canales:2018tbn} and references therein). 
Roughly speaking, scale-dependent gravity extends classical GR solutions after treating the classical coupling as scale-dependent functions, which can be symbolically represented as follows
\begin{align}
\{A_0, B_0, (\cdots)_0\} \rightarrow  \{A_k, B_k, (\cdots)_k\}
\end{align}
Notice that the sub-index $k$ is an arbitrary renormalization scale, which should be connected with one of the 
coordinates of the system.
To account for the relevant interactions, we start by considering a effective action written as
\begin{align} \label{action}
S[g_{\mu \nu},k] \equiv S_{\text{EH}} + S_{\text{M}} + S_{\text{SD}}
\end{align} 
where the terms above mentioned have he usual meaning, namely: 
i)  the Einstein-Hilbert action $S_{\text{EH}}$, 
ii) the matter contribution $S_{\text{M}}$, 
and finally
iii) the scale-dependent term $S_{\text{SD}}$. Moreover, notice that the contribution $S_{\text{M}}$ could account for
either isotropic or anisotropic matter.
For our concrete case, the parameter allowed to vary is Newton’s coupling $G_k$ (or, equivalently, Einstein's coupling $\kappa_k \equiv 8 \pi G_k$).
There are two independent fields, i.e., i) the metric tensor, $g_{\mu \nu}(x)$, and ii) the scale field $k(x)$. 
To obtain the effective Einstein's field equations, we take the variation of \eqref{action} with respect to $g_{\mu \nu}(x)$:
\begin{align}
R_{\mu \nu} - \frac{1}{2}R g_{\mu \nu} = \kappa_k T_{\mu \nu}^{\text{effec}}
\end{align}
In scale-dependent gravity the effective energy-momentum tensor $T_{\mu \nu}^{\text{effec}}$ is defined by
\begin{align}
T_{\mu \nu}^{\text{effec}} \equiv T_{\mu \nu} - \frac{1}{\kappa_k}\Delta t_{\mu \nu}
\end{align}
\begin{equation} \label{Delta_tmunu}
\Delta t_{\mu \nu} \equiv G_k \Bigl(g_{\mu \nu} \Box - \nabla_\mu \nabla_\nu\Bigl) G_k^{-1}
\end{equation}
where the last tensor is obtained after an integration by parts. 

The conventional energy-momentum tensor, $T_{\mu \nu}$, corresponds to matter fields, whereas $\Delta t_{\mu \nu}$ carries the information regarding the running of the gravitational coupling $G_k$. In this sense, when the scale-dependent effect is absent, the aforementioned tensor clearly vanishes. 

In this work, since we are interested in stars with anisotropic matter content, we shall consider an energy-momentum tensor of the form
\begin{equation}
T_{\nu}^{\mu} = \text{diag}\bigl(-\rho, p_r, p_t, p_t\bigl)
\end{equation}
with two different pressures, radial $p_r$, and tangential, $p_t$. Let us comment in passing that in principle one could include the shear as well. It turns out, however, that shear is present only in cases where the metric components depend both on the radial coordinate, $r$ and the time, $t$. This is the case for instance in gravitational collapse, 
see e.g. \cite{Naidu:2005pj}. Since here we are looking for static, spherically symmetric solutions, shear does not contribute and therefore we shall ignore it.

Now, it is essential to improve our comprehension about the running of Newton's coupling, and how such a feature is affected by setting a certain renormalization scale $k$. In this respect, it is well known that General Relativity can be treated as a low energy effective theory and, therefore, it may be viewed as a quantum field theory with an ultraviolet cut-off (see \cite{Dou:1997fg,Donoghue:1994dn} and references therein). That cut-off is parameterized by the renormalization scale $k$, which allows us to surf between a classical and a quantum regime. In order to make progress, the external scale $k$ is usually connected with the radial coordinate.
In the scenario where $G_0 \rightarrow G_k$, the corresponding gap equations are not constant any more. The latter means that non-constant $k = k(x)$ implies that the set of equations of motion does not close consistently. Also, the energy-momentum tensor could be not conserved for a concrete choice of the functional dependence $k = k(x)$. 
This pathology has been analyzed in detail in the context of renormalization group improvement of black holes in asymptotic safety scenarios.
The source of the problem is that a consistency equation is missing, and it can be computed varying the corresponding 
action with respect to the field $k(\cdots)$, i.e., 
\begin{align} \label{Seffec}
\frac{\mathrm{d}}{\mathrm{d}k}S[g_{\mu \nu},k]  &= 0
\end{align}
usually considered to be a variational scale setting procedure \cite{Reuter:2003ca,Koch:2010nn}.
The combination of Eq.~ \eqref{Seffec} with the equations of motion ensures the conservation of the energy-momentum tensor, although an unavoidable problem appears in this approach, i.e., we should know the corresponding $\beta$-functions of the theory. Given that they are not unique, we circumvent the above mentioned computation and, instead of that, we supplement our problem with a auxiliary condition. The energy conditions are four restrictions usually demanded in General Relativity, being the Null Energy Condition (NEC hereafter) the less restrictive of them. We take advantage of this, and we promote the classical coupling to radial-dependent couplings to solve the functions involved. 

Thus, this philosophy of assuring the consistency of the equations by imposing a null energy condition will also be applied for the first time in the following study on interior (anisotropic) solutions of relativistic stars.

\section{Hydrostatic equilibrium of relativistic stars} \label{Hydro}

In this section we briefly review relativistic anisotropic stars in General Relativity and, after that, we will generalize the structure equations in the scale-dependent scenario. Clearly, this work is a natural continuation of our previous work where isotropic relativistic star in the scale-dependent scenario were studied \cite{PRL2}.

The starting point is Einstein's field equations without a cosmological constant
\begin{equation}
G_\nu^\mu = 8 \pi T_\nu^\mu
\end{equation}
where $G_{\mu \nu}$ is Einstein's tensor, and $T_{\mu \nu}$ is the matter stress-energy tensor, which for anisotropic matter takes the form \cite{anisotropia,PRL1}
\begin{equation}
T_\nu ^\mu = \text{diag}(-\rho, p_r, p_t, p_t)
\end{equation}
where $\rho$ is the energy density, $p_r$ is the radial pressure and $p_t$ is the transverse pressure.

Considering a non-rotating, static and spherically symmetric relativistic star in Schwarzschild coordinates, $(t,r,\theta,\phi)$, the most general metric tensor has the form:
\begin{equation}
ds^2 = -e^{\nu} \mathrm{d}t^2 + \frac{1}{1-2m(r)/r} \mathrm{d}r^2 + r^2 \mathrm{d \Omega^2} 
\end{equation}
where we introduce for convenience the mass function $m(r)$, and $\mathrm{d \Omega^2}$ is the line element of the unit two-dimensional sphere. One obtains the Tolman-Oppenheimer-Volkoff equations for a relativistic star with anisotropic matter \cite{anisotropia,PRL1}
\begin{eqnarray}
m'(r)  & = & 4 \pi r^2 \rho(r) \label{mprime} \\
\nu'(r) & = & 2 \frac{m(r)+4 \pi r^3 p_r(r)}{r^2 (1-2m(r)/r)} \\
p_r'(r) & = & -\bigl(p_r(r) + \rho(r) \bigl) \frac{m(r)+4 \pi r^3 p_r(r)}{r^2 (1-2m(r)/r)} + \frac{2 \Delta(r)}{r}  \label{pprime}
\end{eqnarray}
where we define the anisotropic factor $\Delta \equiv p_t - p_r$, and the prime denotes differentiation with respect to the radial coordinate $r$. The special case in which $p_r = p_t$ (i.e., when $\Delta=0$) one recovers the usual Tolman-Oppenheimer-Volkoff equations for isotropic stars \cite{Tolman,OV}.

The exterior solutions is given by the well-known Schwarzschild geometry \cite{SBH}
\begin{equation}
ds^2 = -f(r) \mathrm{d}t^2 + f(r)^{-1} \mathrm{d}r^2 + r^2 \mathrm{d \Omega^2} 
\end{equation}
where $f(r)=1-2M/r$, with $M$ being the mass of the object. Matching the solutions at the surface of the star, the following conditions must be satisfied
\begin{align}
m(r) \Bigl|_{r=R} &= M \\
p_r(r) \Bigl|_{r=R} &= 0 \\
e^{\nu(r)}\Bigl|_{r=R} &= 1 - \frac{2M}{R}
\end{align}
The second condition allows us to compute the radius of the star, the first one allows us to compute the mass of the object, while the last condition allows us to determine the initial condition for $\nu(r)$. Finally, depending on the physics of the matter content the appropriate equation-of-state should be also incorporated, see next section.

Next, we shall now generalize the standard structure equations (valid in General Relativity) in the scale-dependent scenario which accounts for quantum effects. As we already mentioned before, Newton's constant is promoted to a function of the radial coordinate, $G(r)$, and therefore the effective field equations now take the form
\begin{align}
R_{\mu \nu} - \frac{1}{2}R g_{\mu \nu} = 8 \pi G(r) T_{\mu \nu}^{\text{effec}}
\end{align}
where the effective stress-energy tensor has two contributions, namely one from the ordinary matter, $T_{\mu \nu}$,
and another due to the G-varying part, $\Delta t_{\mu \nu}$
\begin{equation}
T_{\mu \nu}^{\text{effec}} = T_{\mu \nu} - \frac{1}{8 \pi G} \: \Delta t_{\mu \nu}
\end{equation}
where the G-varying part was introduced in \ref{Delta_tmunu} (see e.g. \cite{formalism} and references therein for additional details).

Similarly to the classical case, the structure equations valid in the scale-dependent scenario are found to be
\begin{eqnarray}
(G(r) m)' & = & 4 \pi G(r) r^2 \rho^{\text{eff}} \\
\nu'(r) & = & 2 G(r) \: \frac{m + 4 \pi r^3 p_r^{\text{eff}}}{r^2 (1-2G(r)m/r)}
\end{eqnarray}
and we spare the details for the last equation, since it is too long to be shown here. Notice also that when 
$G'(r)=0=G''(r)$ (classical case), the previous set of equations is reduced to the usual TOV equations.

Finally, there is an additional differential equation of second order for $G(r)$, which is the following \cite{formalism}
\begin{equation}
2\frac{G(r)''}{G(r)'} - 4 \frac{G(r)'}{G(r)} = \left[\ln\left(\text{e}^{\nu(r)} \: \frac{1}{1-2m(r)/r}\right) \right]'
\end{equation}
and which must be supplemented by two initial conditions at the center of the star, 
\begin{align}
G(r) \Bigl|_{r=0} &= G_c
\\
G(r)'\Bigl|_{r=0} &= G_1.
\end{align}
%

\section{Interior solutions}

From the formulation of the problem it it clear that there are four equations, namely the three Einstein's field equations plus the additional one for Newton's constant, and six unknown quantities, namely two metric potentials, the r-varying gravitational coupling, and the energy density and the pressures of anisotropic matter. Therefore one is allowed to start
by assuming two conditions. As usual in studies of relativistic stars with anisotropic matter, we shall assume a given
density profile with a reasonable behavior as well as a certain equation-of-state (EoS). Therefore, before we proceed to integrate the structure equations, we must specify the matter source first.

\subsection{Equation-of-state and density profile}

Matter inside the stars is modelled as a relativistic gas of de-confined quarks described by the MIT bag model\cite{bagmodel1,bagmodel2}, where there is a simple analytic function, relating the energy density to the pressure of the fluid, that is
\begin{equation}
p_r = k (\rho - \rho_s)
\end{equation}
where $k$ is a dimensionless numerical factor, while $\rho_s$ is the surface energy density.
The MIT bag model is characterized by 3 parameters, namely i) the QCD coupling constant, $\alpha_c$, ii) the mass of the strange quark, $m_s$, and iii) the bag constant, $B$. The numerical values of $k$ and $\rho_s$ depend on the choice of
$m_s,\alpha_c,B$. In this work we shall consider the extreme model SQSB40, where $m_s=100$~MeV, $\alpha_c=0.6$ and $B=40$~MeV~fm$^{-3}$. In this model $k=0.324$ and $\rho_s=3.0563 \times 10^{14}$~g~cm$^{-3}$ \cite{SQSB40}.

What is more, given that the number of unknown quantities exceeds the number of equations, we may assume a particular density profile $\rho(r)$ as was done for instance in \cite{anisotropia,PRL1}. In our case, we have selected the following density profile:
\begin{align}
\rho &= b \frac{3 + a r^2}{(1 + a r^2)^2}
\end{align}
which is a monotonically decreasing function of the radial coordinate $r$, the central value of which 
is $\rho_c \equiv \rho(0)=3b$. The two free parameters $a,b$ have dimensions $[L]^{-2}$, and they will be taken to be
\begin{eqnarray}
a & = & \frac{\tilde{a}}{(45~km)^2} \\
b & = & \frac{\tilde{b}}{(45~km)^2}
\end{eqnarray}
where now $\tilde{a},\tilde{b}$ are dimensionless numbers.

\subsection{Initial and matching conditions}

Next, since the energy density and the radial pressure are known, we obtain a closed system for $m(r),\nu(r),G(r)$ using the tt and the rr field equations combined with the equation for $G(r)$. Once these are determined, the last field equation allows us to compute the transverse pressure $p_t$ and the anisotropic factor $\Delta$.

Since in this work we assume a vanishing cosmological constant, the exterior solutions is still given by the well-known Schwarzschild geometry, and therefore the matching conditions remains the same as in GR. In the SD scenario there is an additional condition, which requires that Newton's constant must take precisely the classical value at the surface of the star. Therefore, in the SD scenario, the matching conditions are the following:
\begin{align}
m(r) \Bigl|_{r=R} &= M \\
p_r(r) \Bigl|_{r=R} &= 0 \\
e^{\nu(r)}\Bigl|_{r=R} &= 1 - \frac{2M}{R} \\
G(r) \Bigl|_{r=R} &= 1
\end{align}
To integrate the structure equations for $m(r),G(r),\nu(r)$ we impose the initial conditions at the center of the star
\begin{eqnarray}
m(r=0) & = & 0 \\
\nu(r=0) & = & \nu_c \\
G(r=0) & = & G_c \\
G'(r=0) & = & \pm \frac{0.0002}{km}
\end{eqnarray}
where we consider two distinct cases, namely that $G(r)$ can be either a decreasing or an increasing function of $r$, and we fix the absolute value of $G'(r=0)$. The central values $G_c,\nu_c$, in principle unknown, are determined demanding that the matching conditions for $G(R),\nu(r)$, i.e.
\begin{equation}
G(r=R) = 1, \; \; \; \; \; \; \; e^{\nu(r=R)}=1-\frac{2M}{R}
\end{equation}
are satisfied. It should be emphasized here that if $G_c$ and $\nu_c$ are picked up at random, the above matching conditions are not satisfied. Instead, they are satisfied only for the specific initial conditions shown in tables \ref{table:In_Cond_One} and \ref{table:In_Cond_Two}.

\subsection{Numerical results}

\begin{figure*}[ht]
\centering
\includegraphics[width=0.48\textwidth]{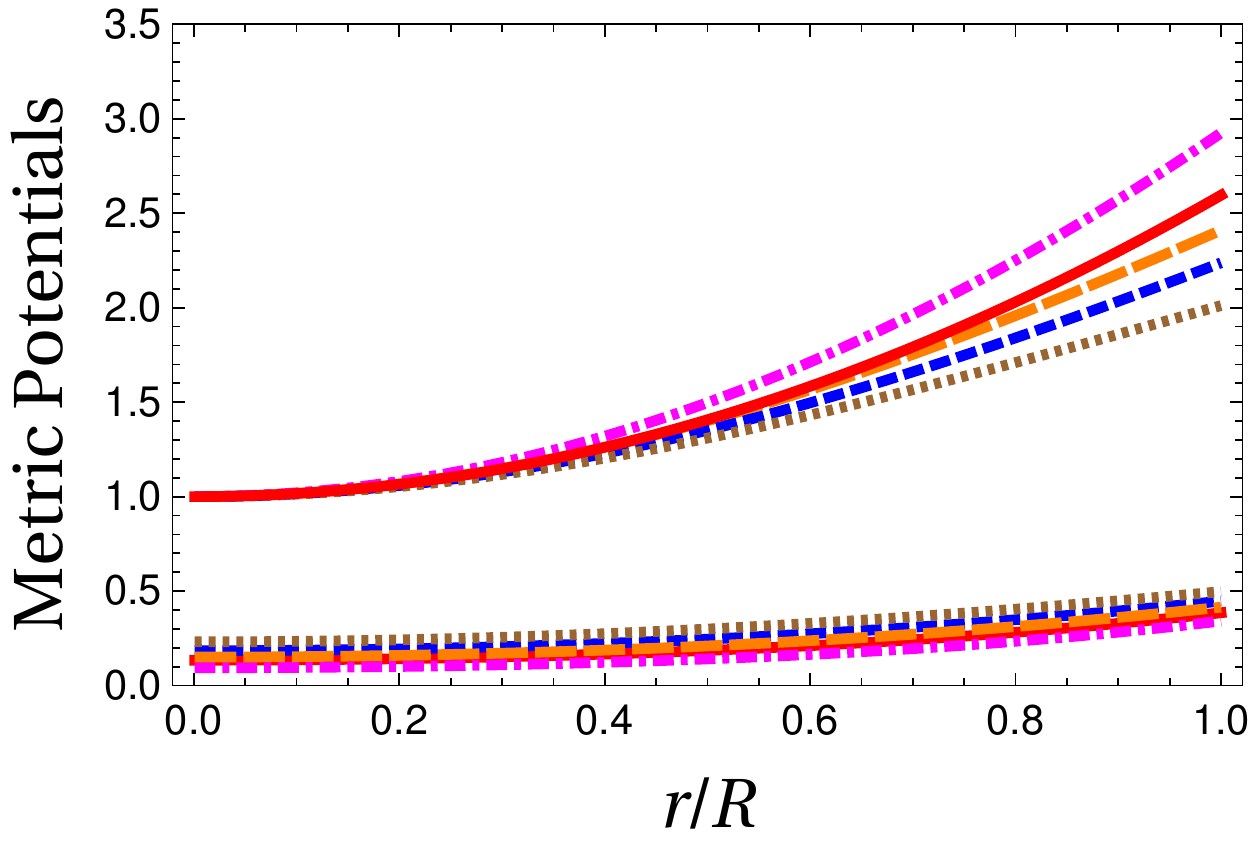}   \
\includegraphics[width=0.48\textwidth]{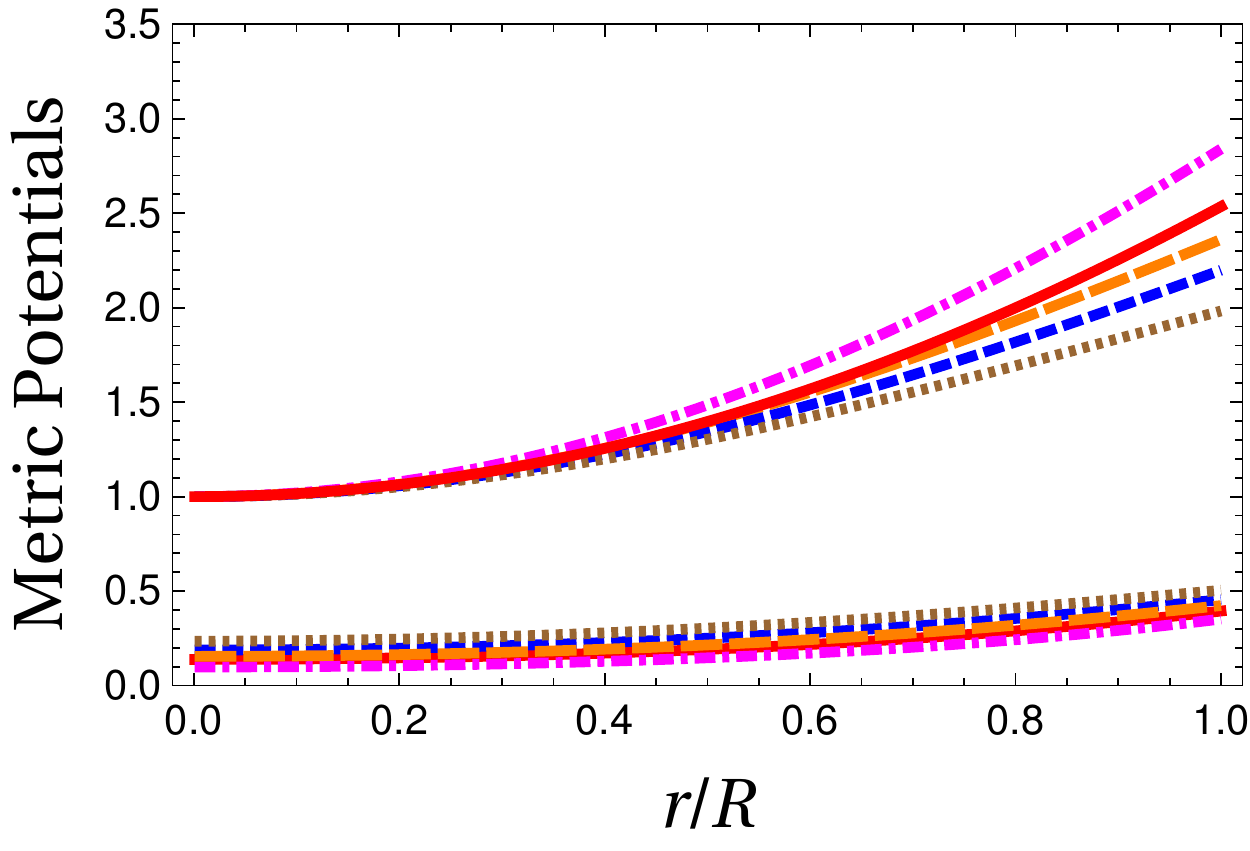}   \
\caption{
Metric potentials $e^\nu$ and $e^\lambda$ vs dimensionless radial coordinate $r/R$ for the five plus five interior solutions obtained here, see Tables I and II for the initial conditions, and Tables III and IV for the properties of the stars. They exhibit the usual behavior of interior solutions of relativistic stars, i.e. they are increasing functions of $r$, and $e^\nu$ always remains below $e^\lambda$. Since $m(r=0)=0$, the metric potential $e^\lambda$ always starts from unity.
{\bf{LEFT:}} Solutions 1-5 corresponding to the case $G'(r=0)=-0.0002/km$ (Table I). 
Shown are: 
  i) Solution 1 (solid red line), 
 ii) Solution 2 (short dashed blue line),
iii) Solution 3 (dotted brown line), 
 iv) Solution 4 (dot-dashed magenta line),
  v) Solution 5 (long dashed orange line).
{\bf{RIGHT:}} Same as left panel, but for solutions 1-5 corresponding to the case $G'(r=0)=+0.0002/km$ (Table II).  
}
\label{fig:1}
\end{figure*}

\begin{figure*}[ht]
\centering
\includegraphics[width=0.48\textwidth]{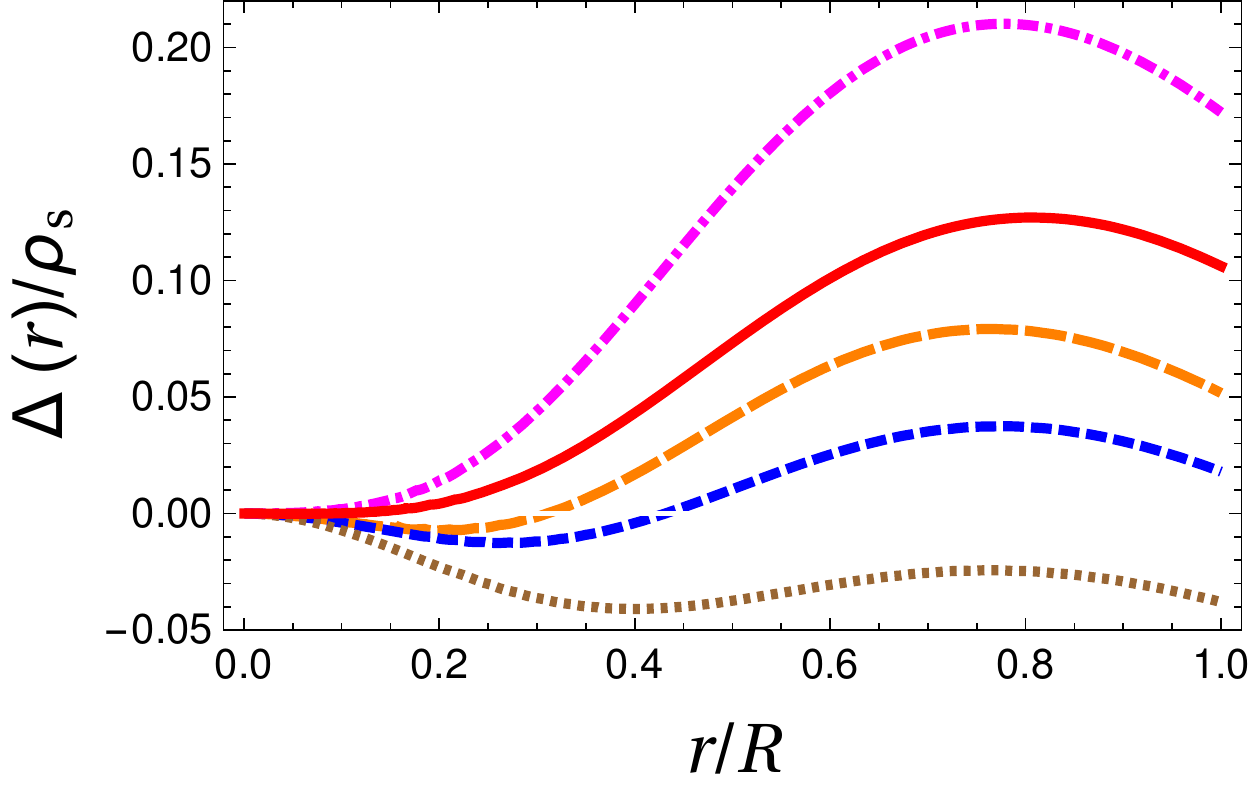}   \
\includegraphics[width=0.48\textwidth]{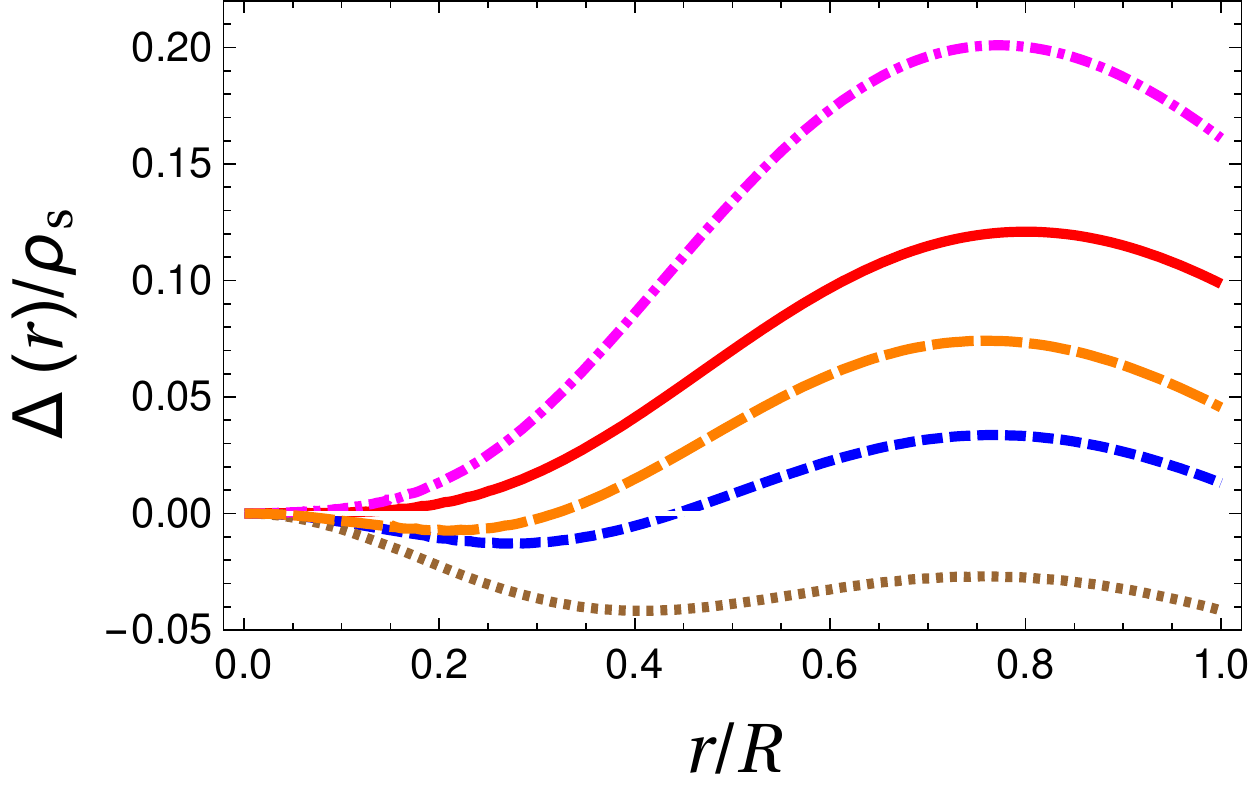}   \
\caption{
Normalized anisotropic factor $\Delta/\rho_s$ vs dimensionless radial coordinate $r/R$ for the five plus five interior solutions obtained here. It starts from zero at the centre of the stars, but it does not have to vanish at the surface.
{\bf{LEFT:}} Solutions 1-5 corresponding to the case $G'(r=0)=-0.0002/km$ (Table I). 
Shown are: 
  i) Solution 1 (solid red line), 
 ii) Solution 2 (short dashed blue line),
iii) Solution 3 (dotted brown line), 
 iv) Solution 4 (dot-dashed magenta line),
  v) Solution 5 (long dashed orange line).
{\bf{RIGHT:}} Same as left panel, but for solutions 1-5 corresponding to the case $G'(r=0)=+0.0002/km$ (Table II).  
}
\label{fig:3}
\end{figure*}

\begin{figure*}[ht]
\centering
\includegraphics[width=0.48\textwidth]{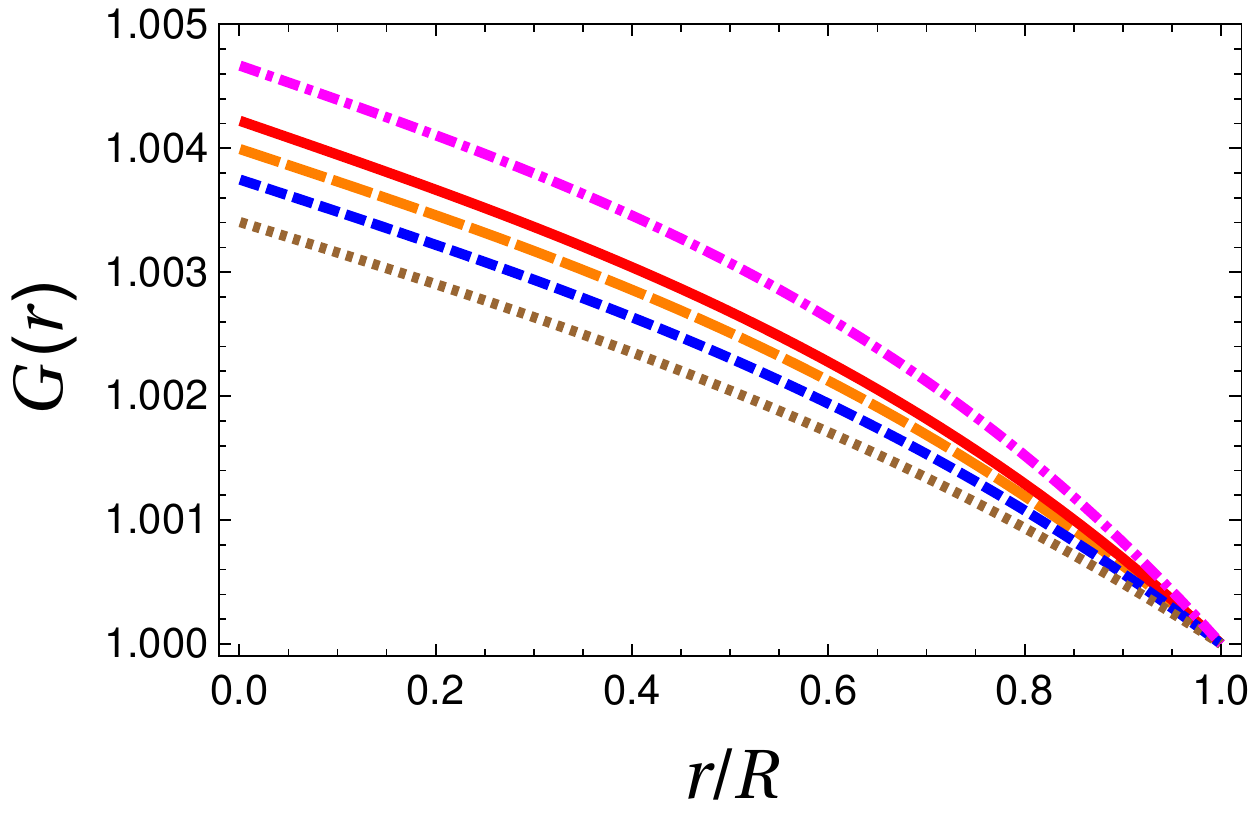}   \
\includegraphics[width=0.48\textwidth]{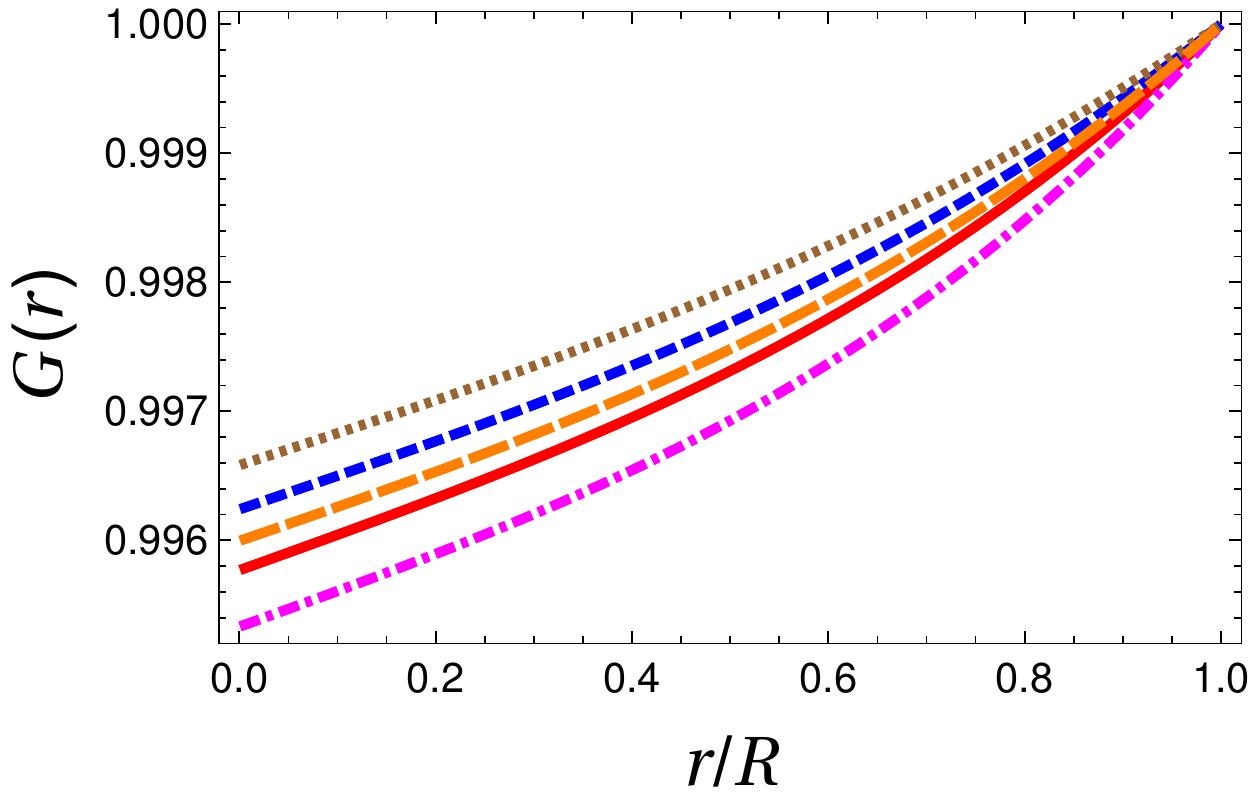}   \
\caption{
Running (r-varying) Newton's constant vs dimensionless radial coordinate $r/R$ for the five plus five interior solutions obtained here.
{\bf{LEFT:}} Solutions 1-5 corresponding to the case $G'(r=0)=-0.0002/km$ (Table I). 
Shown are: 
  i) Solution 1 (solid red line), 
 ii) Solution 2 (short dashed blue line),
iii) Solution 3 (dotted brown line), 
 iv) Solution 4 (dot-dashed magenta line),
  v) Solution 5 (long dashed orange line).
{\bf{RIGHT:}} Same as left panel, but for solutions 1-5 corresponding to the case $G'(r=0)=+0.0002/km$ (Table II).  
}
\label{fig:4}
\end{figure*}

\begin{figure}[ht]
\centering
\includegraphics[width=0.48\textwidth]{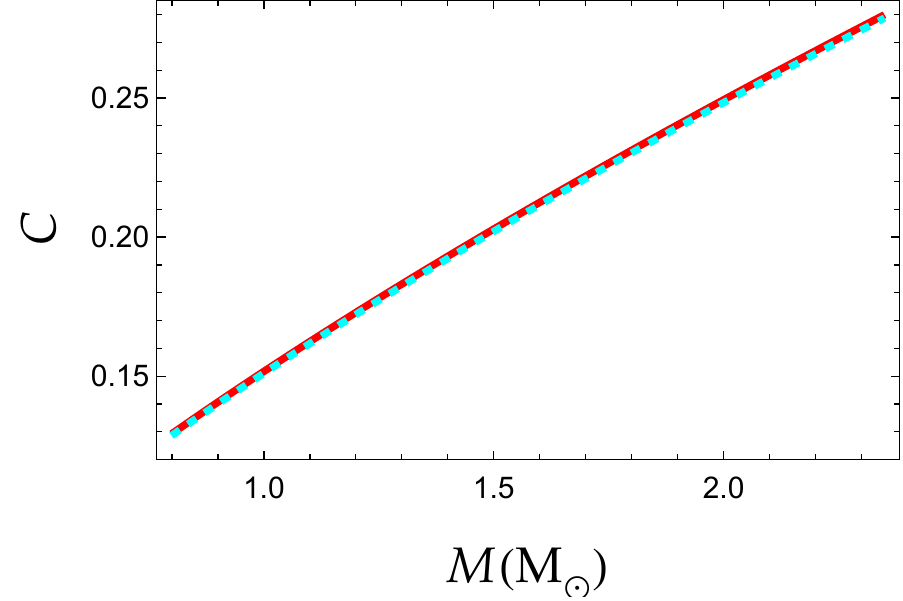}   \
\includegraphics[width=0.48\textwidth]{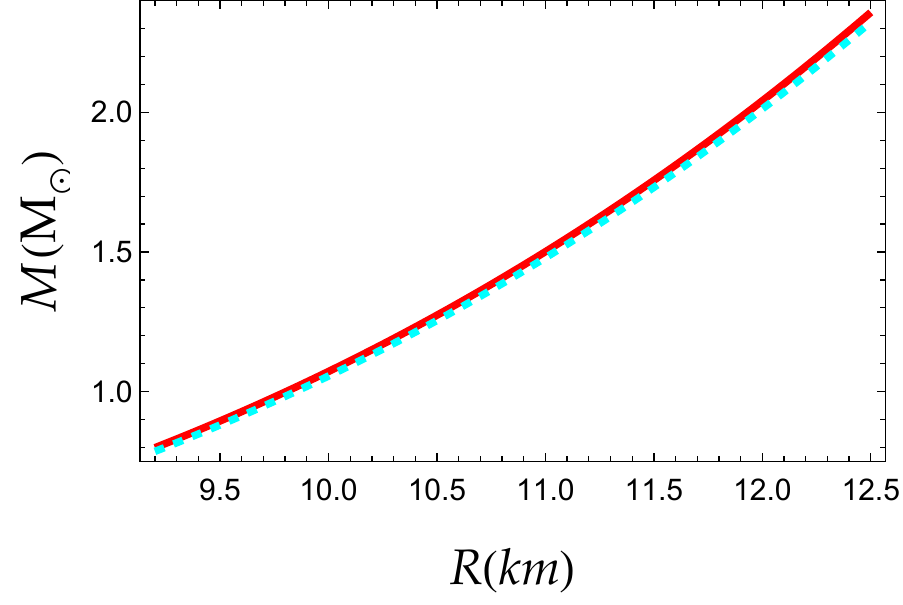}   \
\includegraphics[width=0.48\textwidth]{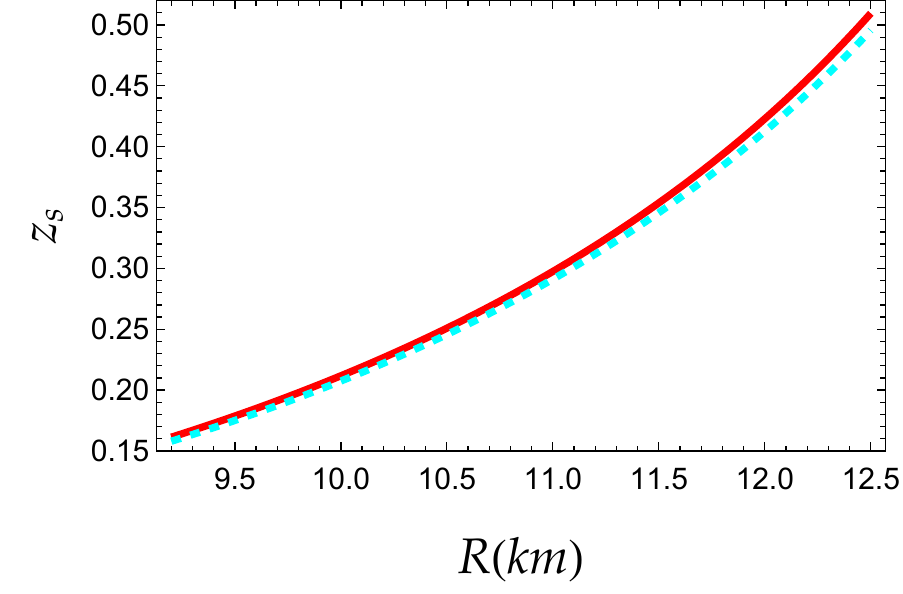}   \
\caption{
{\bf{TOP:}} Factor of compactness, $C$, versus mass, $M$ (in solar masses). Solid red line corresponds to the case $G'(r=0)=-0.0002/km$, while dashed cyan line corresponds to the case $G'(r=0)=+0.0002/km$. 
{\bf{MIDDLE:}} Mass-to-radio profiles for the two cases regarding the sign of $G'(r=0)$. The color code is the 
same as in the top panel.
{\bf{BOTTOM:}} Surface red-shift, $z_s$, (see text) versus radius $R$ (in km). The color code is the same as in the top
panel.
}
\label{fig:6}
\end{figure}

Our main numerical results are summarized in the tables and in the figures below. 
In particular, first we present in detail five representative solutions for $G'(r=0) < 0$, and five more for $G'(r=0) > 0$. The numerical values of $\tilde{a},\tilde{b}$ are shown as well. In Tables~\ref{table:In_Cond_One} and \ref{table:In_Cond_Two} we show the initial conditions for $G(r)$ and $\nu(r)$ for a negative and a positive $G'(r=0)$, respectively, while in Tables~\ref{table:First_set} and \ref{table:Second_set} we show the properties (i.e. mass, radius and compactness) of strange quark stars for positive and negative $G'(r=0)$, respectively. Our results show that for a given density profile (given pair $a,b$ and given radius), a decreasing Newton's constant (Table \ref{table:Second_set}) implies a more massive star and consequently a higher compactness factor in comparison with an increasing Newton's constant (Table \ref{table:First_set}). The mass-to-radius profiles as well as the factor of compactness versus the mass, and the surface
red-shift, $z_s$, versus the radius of the objects are shown in the three panels of Fig.~\ref{fig:6}. The surface red-shift, an important quantity to astronomers, is given by \cite{Red1,Red2,Red3}
\begin{equation}
z_s = -1 + \left( 1 - 2 \frac{M}{R} \right)^{-1/2}
\end{equation}
Fig.~\ref{fig:1} shows the metric potentials $e^{\nu(r)}$ and $e^{\lambda(r)}$ as a function of the dimensionless radial coordinate $r/R$. Fig.~\ref{fig:2} shows the (normalized) energy density as well as the radial and transverse pressure versus $r/R$ for the five plus five cases considered here, while Fig.~\ref{fig:3} shows the anisotropic factor $\Delta/\rho_s$ vs $r/R$. Fig.~\ref{fig:4} shows the scale-dependent gravitational coupling as a function of radial coordinate. 
All solutions, irrespectively of the sign of $G'(r)$, are found to be well behaved, realistic solutions,, which tend to
$G(r=R) = 1$ at the surface of the star, which was imposed right from the start.

\subsection{Stability and energy conditions}

The interior solutions obtained here must be able to describe realistic astrophysical configurations. In this subsection we check if stability criteria as well as the energy conditions are fulfilled or not. First, regarding stability, we impose the condition $\Gamma > 4/3$ \cite{bondi64,Chandra,Moustakidis}, where the adiabatic index $\Gamma$ is defined by
\begin{equation}
\Gamma \equiv c_s^2 \left[ 1 + \frac{\rho}{p_r} \right]
\end{equation}
with $c_s$ being the sound speed defined by
\begin{equation}
c_s^2 \equiv \frac{dp_r}{d \rho}
\end{equation}
For the linear EoS considered here, the speed of sound is a constant, $c_s^2 = k$. 

\begin{figure*}[ht]
\centering
\includegraphics[width=0.48\textwidth]{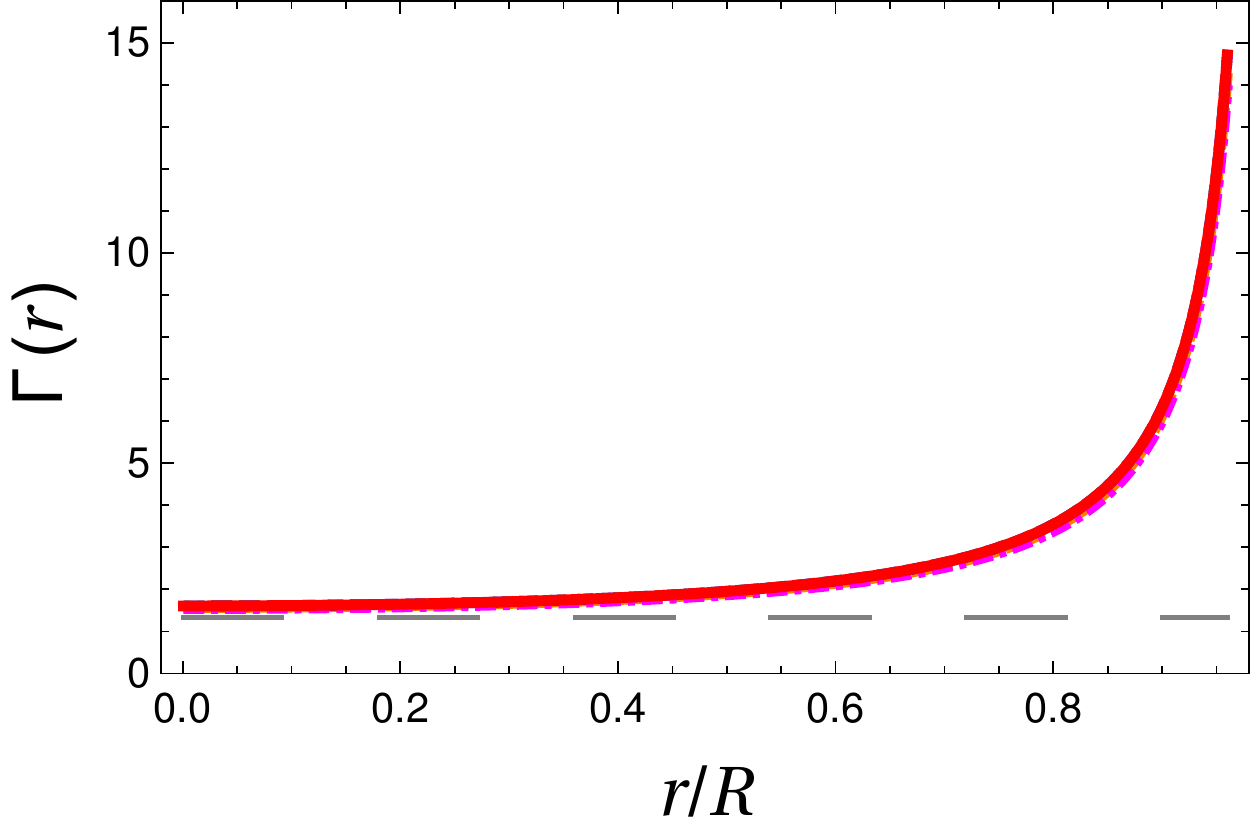}   \
\includegraphics[width=0.48\textwidth]{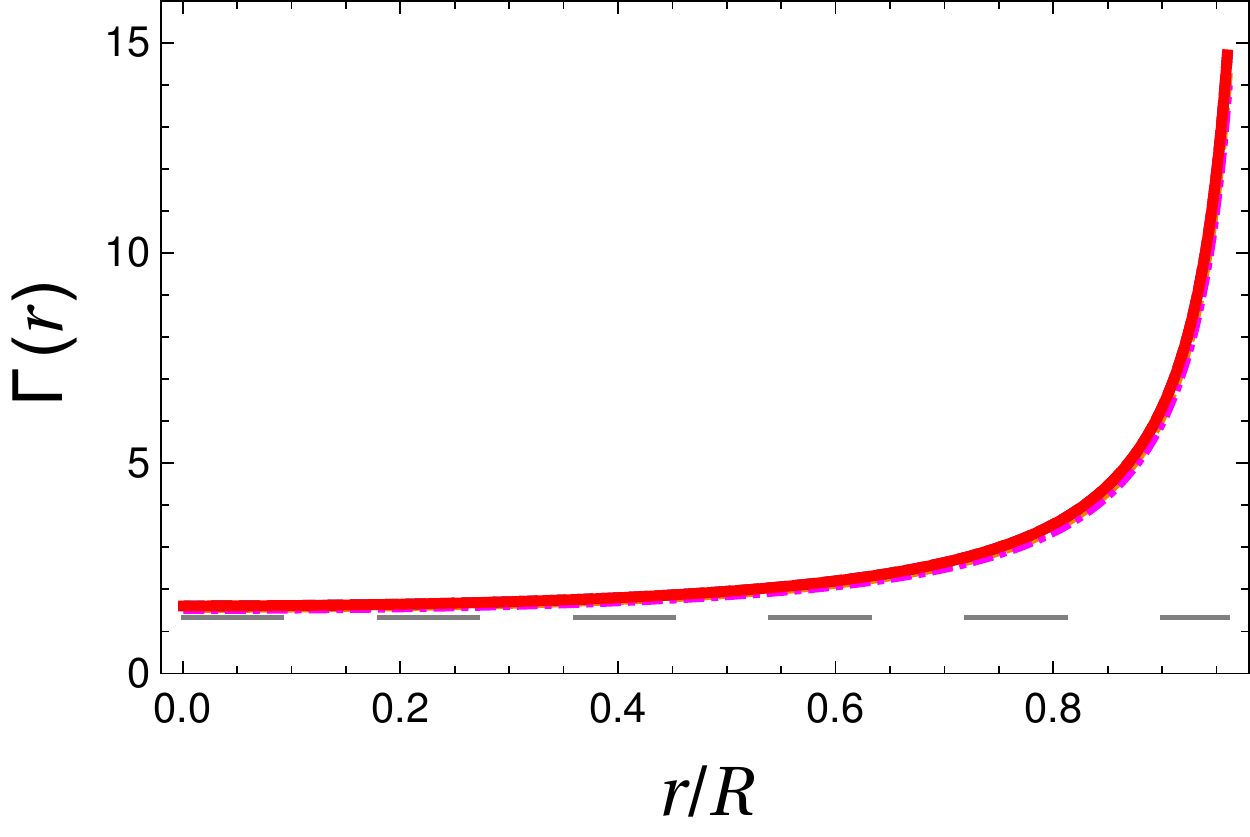}   \
\caption{
Adiabatic index $\Gamma$ versus dimensionless radial coordinate $r/R$ for the 
five plus five interior solutions obtained here. The horizontal dashed line corresponds to the Newtonian limit $4/3$.
{\bf{LEFT:}} Solutions 1-5 corresponding to the case $G'(r=0)=-0.0002/km$ (Table I). 
Shown are: 
  i) Solution 1 (solid red line), 
 ii) Solution 2 (short dashed blue line),
iii) Solution 3 (dotted brown line), 
 iv) Solution 4 (dot-dashed magenta line),
  v) Solution 5 (long dashed orange line).
{\bf{RIGHT:}} Same as left panel, but for the solutions 1-5 corresponding to the case $G'(r=0)=+0.0002/km$ (Table II). 
}
\label{fig:5}
\end{figure*}

Fig.~\ref{fig:5} shows that $\Gamma > 4/3$ for all the models considered here, both for positive (left panel) and negative $G'(r=0)$ (right panel).

\begin{figure*}[ht]
\centering
\includegraphics[width=0.48\textwidth]{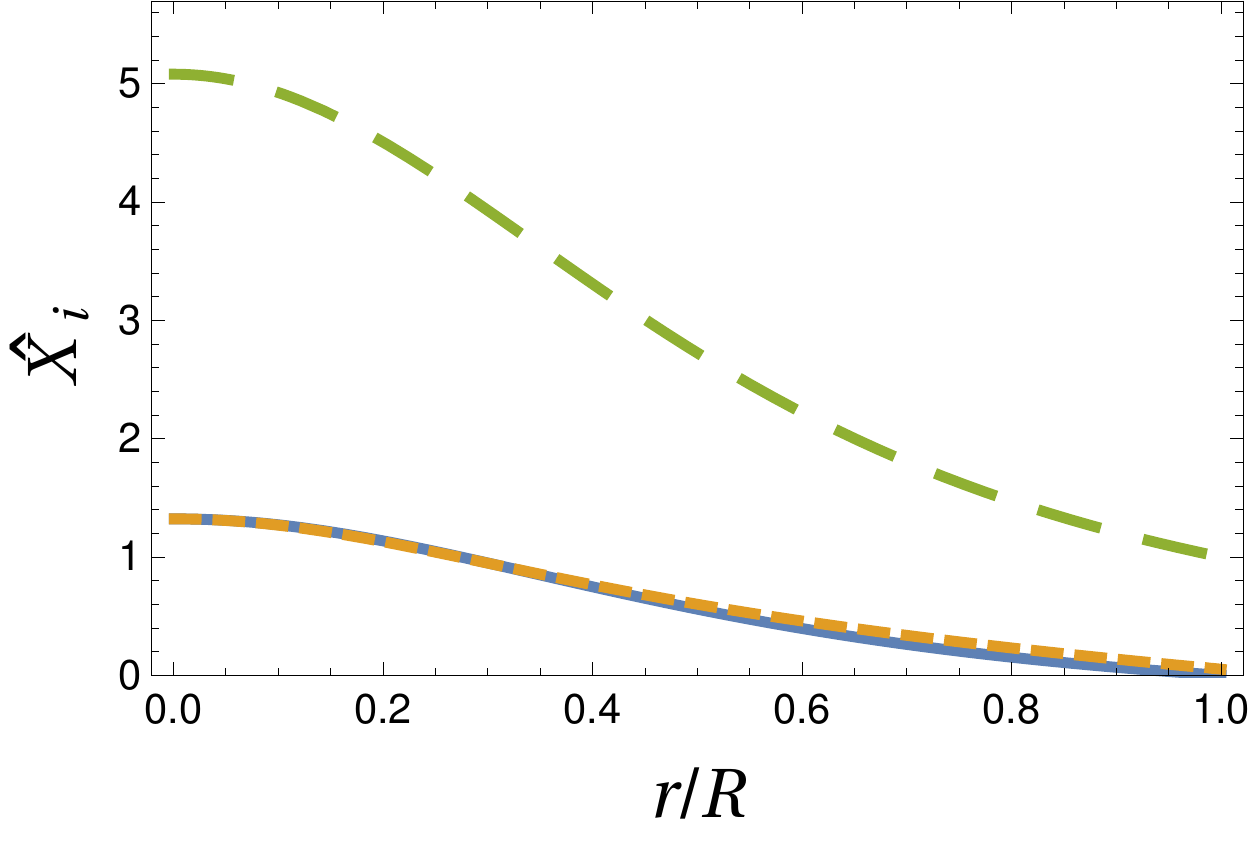}   \
\includegraphics[width=0.48\textwidth]{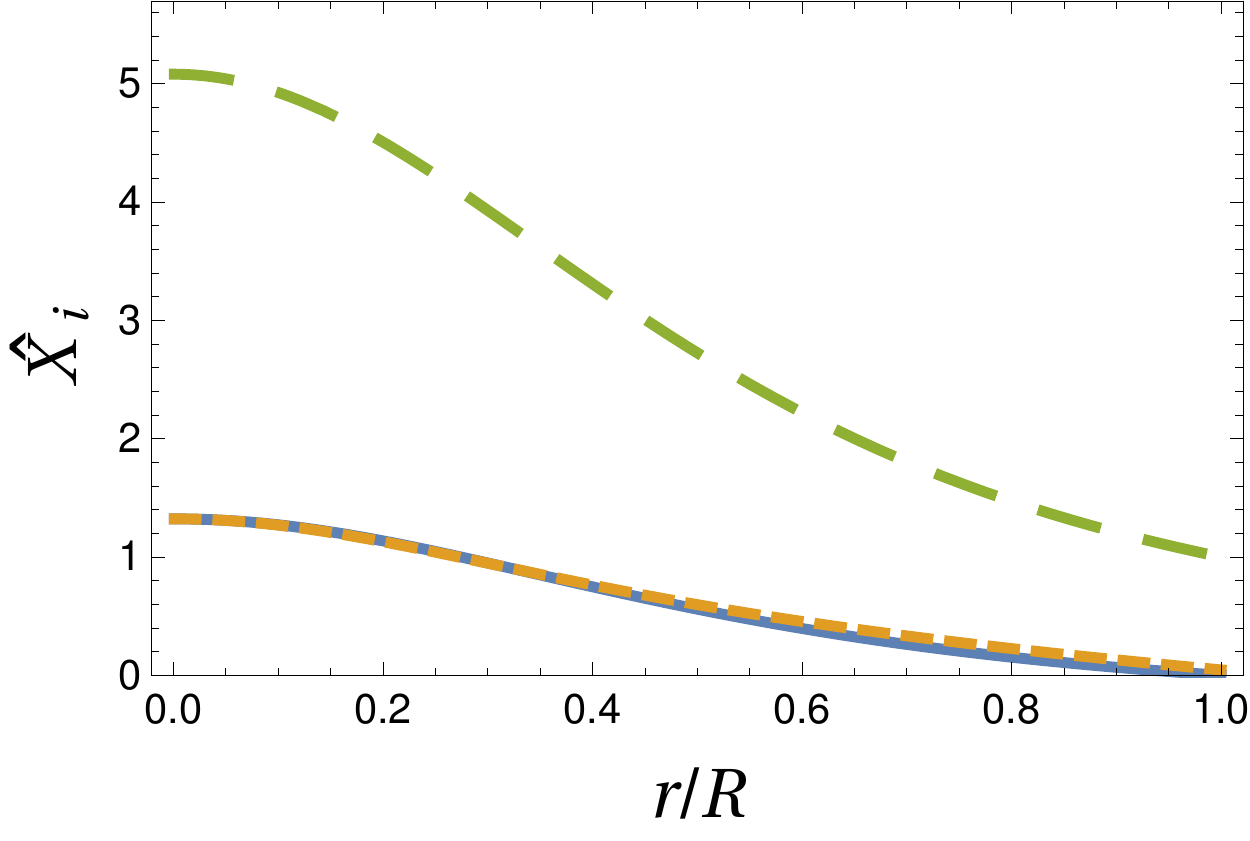}   \
\caption{
Dimensionless energy density and pressures (both radial and tangential) 
$X_i \equiv (\rho,p_r,p_t)/\rho_s$ vs dimensionless radial coordinate $r/R$ 
for the 5th solution obtained here.
{\bf{LEFT:}} Solution for the case $G'(r=0)=-0.0002/km$ (5th solution in Table I). 
{\bf{RIGHT:}} Solution for the case $G'(r=0)=+0.0002/km$ (5th solution in Table II ). 
Shown are: 
  i) dimensionless radial pressure $p_r/\rho_s$ (solid blue line),
 ii) dimensionless transverse pressure $p_t/\rho_s$ (short dashed orange line),
iii) dimensionless density $\rho/\rho_s$ (long dashed green line). 
The energy density starts from its central value, $\rho_c$, and it monotonically decreases until it reaches its surface value, $\rho_s$. The pressures start from the same value at the centre of the stars (which implies that the anisotropy factor vanishes there, see next figure), and they monotonically decrease until $p_r$ vanishes at the surface, whereas $p_t$ does not have to vanish.
}
\label{fig:2}
\end{figure*}

Next, regarding energy conditions, we require that \cite{Ref_Extra_1,Ref_Extra_2,Ref_Extra_3,ultimo,PR3}
\begin{equation}
\mbox{WEC:} \,\,\, \rho \geq 0\,, \,\,\, \rho + p_{r,t} \geq 0\,,
\end{equation}
\begin{equation}
\mbox{NEC:} \,\,\, \rho + p_{r,t}  \geq  0\,,
\end{equation}
\begin{equation}
\mbox{DEC:} \,\,\, \rho \geq \lvert p_{r,t} \rvert\,,
\end{equation}
\begin{equation}
\mbox{SEC:} \,\,\, \rho + p_{r,t}  \geq  0\,, \,\,\, \rho + p_r + 2 p_t \geq 0\,.
\end{equation} 
According to the interior solution shown in Fig.~ \ref{fig:2}, we observe that i) all three quantities $\rho,p_r,p_t$ are positive throughout the star, and ii) the energy density always remains larger that both $p_r,p_t$. Clearly all energy conditions are fulfilled. Hence, we conclude that the interior solutions found here are well behaved solutions within scale-dependent gravity, capable of describing realistic astrophysical configurations.

As a final remark it should be stated here that in the present article we took a modest step towards the investigation of spherically symmetric, anisotropic strange quark stars in the scale-dependent scenario assuming the MIT bag model equation-of-state. Regarding future work, one may consider i) more sophisticated equations-of-state for quark matter \cite{refine1,refine2,refine3}, ii) rotating stars, or iii) other types of compact objects, such as neutron stars or white dwarfs. It would be interesting to study those issues in scale-dependent gravity in forthcoming articles.


\begin{table*}
\centering
\caption{Initial conditions for five interior solutions for $G'(r=0)=-0.0002/km$.}
\begin{tabular}{ccc}
\hline
$No$ of solution &  $G_c$ & $\nu_c$   \\
\hline
1      & 1.00422 &  -2.00721  \\ 
\hline
2      & 1.00375 &  -1.69046  \\
\hline
3      & 1.00341 &  -1.46346  \\
\hline
4      & 1.00467 &  -2.36200   \\
\hline
5      & 1.00400 &  -1.89206  \\
\hline
\end{tabular}
\label{table:In_Cond_One}
\end{table*}


\begin{table*}
\centering
\caption{Initial conditions for five interior solutions for $G'(r=0)=0.0002/km$.}
\begin{tabular}{ccc}
\hline
$No$ of solution &  $G_c$ & $\nu_c$   \\
\hline
1      & 0.99577 &  -1.97771  \\ 
\hline
2      & 0.99624 &  -1.67030  \\
\hline
3      & 0.99658 &  -1.44859  \\
\hline
4      & 0.99533 &  -2.32040   \\
\hline
5      & 0.99599 &  -1.86682  \\
\hline
\end{tabular}
\label{table:In_Cond_Two}
\end{table*}


\begin{table*}
\centering
\caption{Properties of five interior solutions assuming a positive $G'(r=0)$.}
\begin{tabular}{cccccc}
\hline
$No$ of solution &  $R[km]$ & $M[M_{\odot}]$ & $C=M/R$ & $\tilde{a}$ & $\tilde{b}$  \\
\hline
1      & 13.67  & 2.783  & 0.303  &  18   &  0.7   \\ 
\hline
2      & 12.97  & 2.377  & 0.273  &  20   &  0.7   \\
\hline
3      & 12.37  & 2.060  & 0.248  &  22   &  0.7   \\
\hline
4      & 13.75  & 2.995  & 0.324  &  22   &  0.85   \\
\hline
5      & 13.13  & 2.544  & 0.288  &  22   &  0.78   \\
\hline
\end{tabular}
\label{table:First_set}
\end{table*}


\begin{table*}
\centering
\caption{Properties of five interior solutions assuming a negative $G'(r=0)$.}
\begin{tabular}{cccccc}
\hline
$No$ of solution &  $R[km]$ & $M[M_{\odot}]$ & $C=M/R$ & $\tilde{a}$ & $\tilde{b}$  \\
\hline
1      & 13.67  & 2.825  & 0.307  &  18   &  0.7   \\ 
\hline
2      & 12.97  & 2.412  & 0.277  &  20   &  0.7   \\
\hline
3      & 12.37  & 2.090  & 0.251  &  22   &  0.7   \\
\hline
4      & 13.75  & 3.042  & 0.329  &  22   &  0.85   \\
\hline
5      & 13.13  & 2.582  & 0.293  &  22   &  0.78   \\
\hline
\end{tabular}
\label{table:Second_set}
\end{table*}



\section{Conclusions}

Summarizing our work, in the present article we have obtained well behaved 
interior solutions for relativistic stars with anisotropic matter in the 
scale-dependence scenario. In particular, we have investigated the properties of strange stars 
adopting the extreme SQSB40 MIT bag model, and assuming for quark matter a linear EoS. 
First we derived the generalized structure equations describing the hydrostatic equilibrium 
of the stars for a non-vanishing anisotropic factor. Those new equations generalize the 
usual TOV equations valid in GR, which are recovered when Newton's constant 
is taken to be a constant, $G'(r)=0=G''(r)$. Next, assuming a certain profile for the 
energy density, we numerically integrated the structure equations for the system $m(r),\nu(r),G(r)$, and we 
computed the radius, the mass as well as the factor of compactness of the 
stars for a varying Newton's constant, either increasing or decreasing, throughout the 
objects. Moreover, we have shown that the energy conditions are fulfilled, and that 
the Bondi's stability condition, $\Gamma > 4/3$, is satisfied as well. In both cases, 
$G'(r=0) > 0$ and $G'(r=0) < 0$, we obtained well behaved solutions describing realistic astrophysical configurations, although a decreasing Newton's constant throughout the 
objects leads to slightly more massive and more compact stars.


\section*{Acknowlegements}

We wish to thank the anonymous reviewer for valuable comments and suggestions.
The authors G.~P. and I.~L. thank the Fun\-da\c c\~ao para a Ci\^encia e Tecnologia 
(FCT), Portugal, for the financial support to the Center for Astrophysics and 
Gravitation-CENTRA, Instituto Superior T\'ecnico, Universidade de Lisboa, through 
the Project No.~UIDB/00099/2020 and No.~PTDC/FIS-AST/28920/2017. The author A.~R. 
acknowledges DI-VRIEA for financial support through Proyecto Postdoctorado 2019 VRIEA-PUCV.


\end{document}